\documentclass[fleqn,usenatbib]{mnras}
\usepackage{fix-cm}
\usepackage{newtxtext,newtxmath}

\usepackage[T1]{fontenc}

\DeclareRobustCommand{\VAN}[3]{#2}
\let\VANthebibliography\thebibliography
\def\thebibliography{\DeclareRobustCommand{\VAN}[3]{##3}\VANthebibliography}

\usepackage{graphicx}	
\usepackage{amsmath}	
\usepackage{xspace}

\newcommand{\pc}{\,\mathrm{pc}}
\newcommand{\Msun}{\,\mathrm{M}_{\odot}}
\newcommand{\MsunYr}{\,\mathrm{M}_{\odot}\mathrm{yr}^{-1}}
\newcommand{\Myr}{\,\mathrm{Myr}}

\newcommand{\K}{\,\mathrm{K}}

\newcommand{\kms}{\,\mathrm{km\,s}^{-1}}
\newcommand{\gocm}{\,\mathrm{g\,cm}^{-3}}

\newcommand{\mg}{\,\mathrm{\mu\,G}}

\newcommand{\RHD}{{RHD}\xspace}
\newcommand{\HD}{{HD}\xspace}
\newcommand{\BxWRT}{{BX0.5RT}\xspace}
\newcommand{\BxSRT}{{BX5RT}\xspace}
\newcommand{\BxSSRT}{{BX50RT}\xspace}

\newcommand{\BxWNoRT}{{BX0.5NoRT}\xspace}
\newcommand{\BxSNoRT}{{BX5NoRT}\xspace}
\newcommand{\BxSSNoRT}{{BX50NoRT}\xspace}

\newcommand{\ByWRT}{{\sc BY0.5RT}\xspace}
\newcommand{\BySRT}{{\sc BY5RT}\xspace}
\newcommand{\BySSRT}{{\sc BY50RT}\xspace}

\newcommand{\NONRT}{{\sc NoRT}\xspace}
\newcommand{\RT}{{\sc RT}\xspace}

\newcommand{\ramses}{{\sc Ramses}\xspace}
\newcommand{\ramsesrt}{{\sc Ramses-rt}\xspace}

\defcitealias{Yaghoobi2024}{Paper\,I}

\newcommand{\secref}[1]{Section~\ref{#1}}
\newcommand{\figref}[1]{Figure~\ref{#1}}
\newcommand{\tabref}[1]{Table~\ref{#1}}

\usepackage[normalem]{ulem}
\usepackage{xcolor}

\title[The Influence of Magnetic Fields on SG Star Formation]{The Influence of Magnetic Fields on Second-Generation Star Formation in Globular Clusters}

\author[A. Yaghoobi et al.]{
	A. Yaghoobi, $^{1,2}$\thanks{E-mail:a.yaghoobi@ipm.ir}
	F. Tabatabaei$^{1}$, 
	J. Rosdahl$^{2}$,
	B. Commercon$^{2}$,
	S. Sheikhnezami$^{3}$,
	and  F. Calura$^{4}$
	\\
	$^{1}$ Institute for Research in Fundamental Sciences (IPM), Larak Garden, 19395-5531 Tehran, Iran \\
	$^{2}$ Univ Lyon, Univ Lyon1, Ens de Lyon, CNRS, Centre de Recherche Astrophysique de Lyon UMR5574, F-69230, Saint-Genis-Laval, France\\
	$^{3}$ Department of Physics, Institute for Advanced Studies in Basic Sciences (IASBS), 444 Prof. Yousef Sobouti Blvd., 45137-66731, Zanjan, Iran \\
	$^{4}$ INAF - OAS, Osservatorio di Astrofisica e Scienza dello Spazio di Bologna, via Gobetti 93/3, I-40129 Bologna, Italy
}

\date{Accepted XXX. Received YYY; in original form ZZZ}

\begin{document}
	\label{firstpage}
	\pagerange{\pageref{firstpage}--\pageref{lastpage}}
	\maketitle
	
	\begin{abstract}
		We investigate the previously unexplored role of magnetic fields in the formation of second-generation (SG) stars in proto-globular clusters (GCs) using 3D radiation-magnetohydrodynamical simulations. This study is based on the asymptotic giant branch (AGB) scenario and incorporates photoionization feedback and stellar winds from AGB stars. We model SG formation within a young ($34\Myr$) massive ($10^6\Msun$) proto-GC moving through a magnetized, homogeneous interstellar medium. Our results indicate that variations in magnetic field strength and orientation significantly influence the gas geometry and SG star-forming regions around the cluster. Overall, magnetic fields limit SG formation to the very center of the cluster, with stronger magnetic fields tending to form more compact SG clusters.
		For magnetic field strengths of $0.5$ and $5\mg$, we observe no substantial changes in the mass of formed SG stars. However, with a strong $50\mg$ field, we see a $25$ percent increase or a $70$ percent decrease in total SG mass, for a field aligned parallel or perpendicular to the cluster's motion, respectively. This variation reflects how magnetic fields influence gas accretion, as our results suggest that gas accreted from the interstellar medium (ISM) slightly dominates over AGB ejecta in the cluster — except in cases of strong perpendicular fields, where gas accretion is efficiently suppressed.
		Additionally, stronger magnetic fields limit the cluster's ability to retain its ejecta, leading to the formation of stars with lower helium abundances. On the other hand, a strong perpendicular magnetic field produces SG stars that originate from AGB ejecta and exhibit the highest helium abundances. 
	\end{abstract}
	
	
	
	
	\section{Introduction}
	One of the open questions in modern Astrophysics is the formation of multiple stellar populations (MSPs) in individual globular clusters (GCs). 
	These populations show anomalous variations in light element abundances, i.e., He, C, N, O, Na, Mg, and Al \citep{Sneden1992,Gratton2004,carretta2009,carretta2010, Gratton2012}. The stars within GCs with the same chemical compositions as those in field stars are known as the first population or first generation (FG).
	The other GC stars with anomalous composition are generally referred to as the second population or second generation \citep[SG,][]{bastian2018,milone2022}. The SG stars are enriched in N and Na and depleted in C and O, whereas the sum of the C, N, and O abundances inside the stars belonging to both populations is interestingly observed to be generally constant \citep{Dickens1991,minniti1993,carretta2009,gratton2013,Yong2015}. These variations are observed in almost all massive globular clusters younger than $2 {\rm\, Gyr}$ \citep{bastian2018}, but not within young massive clusters (YMCs; with an age below $100\Myr$). This age dependence strongly suggests that multiple population formation occurred preferentially during the epoch of peak star formation at redshift $z > 2$ \citep{D'Ercole2016}.
	Recent results from the Hubble Space Telescope (HST) photometry reveal that the observed helium (He) spread (${\rm \Delta} Y$) of SG stars in GCs strongly correlates with their present-day and initial masses. The number ratio of SG to FG stars also correlates with cluster mass, with more massive clusters having larger helium spreads and SG number fractions \citep{milone2017,Lagioia2019,milone2020}.
	
	Such an abundance pattern is expected from the yields of hot hydrogen burning in intermediate-mass \citep[4-8$\Msun$,][]{Maeder2006} and massive stars \citep[$ m \ge 15 \Msun$,][]{Karakas2014}. Consequently, the majority of scenarios suggested for the formation of MSPs propose that the ejected material from FG stars is the origin of anomalous compositions seen in MSPs \citep{D'Ercole2008,Mink2009,krause2013,Denissenkov2014}.
	
	The AGB scenario is one of the main models proposed for the formation of MSPs in globular clusters. This model \citep{DAntona1983,Renzini1988,D'Ercole2008, D'Ercole2010,D'Ercole2016, Bekki2017} suggests that the MSPs are formed sequentially from the ejecta released by AGB stars and the pristine gas left over from the formation of the FG stars in the first few hundred million years of the cluster formation. This model has several essential requirements:
	\begin{enumerate}
		\item Supernova (SN) ejecta should not significantly contribute to the SG formation process, as they would result in variations in heavy element abundances that do not match observations. Previous studies \citep[e.g.][]{calura2015} have shown that SN ejecta can efficiently escape the cluster potential due to their high velocities. However, it is possible that extremely massive clusters could retain their SN ejecta and exhibit some variations in heavy elements, as observed in some massive GCs such as Ruprecht 106 \citep{Frelijj2021}.
		
		\item The cluster must retain its AGB ejecta to provide the enriched material necessary for SG formation. The typical velocity dispersion of proto-globular clusters is similar (in order of magnitude) to the wind velocity \citep{naiman2018,Yaghoobi2022a}, allowing massive proto-GCs to retain their AGB stellar winds effectively.
		
		\item The cluster must accrete pristine gas to dilute the AGB ejecta. The light element abundances of AGB stellar winds do not match the observed SG abundances, which lie between the FG stars and AGB material abundances. Therefore, the AGB material is proposed to be diluted by the pristine gas having the same abundances as the FG population \citep{D'Ercole2016}. However, it has been computed that the amount of required pristine gas cannot exceed $10$ percent of the FG mass, as larger fractions would result in abundance patterns that contrast with observations \citep{D'Ercole2010,D'Ercole2011}.
		\item The cluster must lose approximately $90$ percent of its FG stars during long-term dynamical evolution of the cluster to solve the \textit{mass budget problem}. If the total amount of AGB ejecta from a typical cluster, along with the pristine gas required for dilution, is converted into SG stars, the resulting number ratio of SG stars to total stars is lower than the observed ratios, which range from $0.1$ to $0.9$ \citep{milone2020}. Given that the SG stars are more concentrated toward the center than the FG stars, various studies \citep[e.g,][]{Vesperini2021,Sollima2024} have shown that the long-term dynamical evolution of a cluster results in a consistent loss of FG stars, leading to an increase in the SG fraction in GCs.
	\end{enumerate}
	
	These requirements must be thoroughly evaluated under various conditions to assess whether this scenario can effectively form a new generation of stars within a globular cluster and reproduce the present-day properties of MSPs \citep{D'Ercole2008,Conroy2011, Bekki2017b,Bekki2017, calura19}. 
	Using 1D hydrodynamic simulations, \citet{D'Ercole2008} showed that massive clusters can retain the ejecta from AGB stars and form a new generation of stars. It has also been shown that these massive clusters can accrete ambient gas from the interstellar medium (ISM) \citep[][]{lin2007, naiman2011}. The ram pressure stripping exerted by the ISM on a moving cluster can overcome the cluster's gravitational pull in low-mass clusters, resulting in a lower SG mass ($\sim 10^3 \Msun$) compared to the SG mass ($\gtrsim 10^5 \Msun$) formed in massive clusters  \citep{calura19,Yaghoobi2022a}. The effects of Type Ia supernovae explosions are investigated in \citet{lacchin21}. They find that SN explosions weakly affect the SG formation in high-density environments, but strongly at lower densities. Their result for high-density runs shows an iron spread of  $\sim0.14$ dex, which is consistent with the variation found in about $20$ percent of Galactic GCs \citep{Milone2016,Marino2019}. Moreover, we showed in \citet[][]{Yaghoobi2022b,Yaghoobi2024} that ionizing radiation has a key role in SG formation, depending on the cluster mass and ISM density. In low-mass clusters and lower-density environments, the radiation completely suppresses the formation of SG stars. Hence, photoionizing feedback limits SG star formation to massive clusters formed in dense environments. We also showed that in low-density models, the gas content is negligible, matching the minimal gas levels observed in present-day YMCs. These systems, forming in relatively sparse environments compared to the high-density conditions characteristic of high-redshift cluster formation, cannot effectively accumulate either pristine gas or AGB ejecta in their central regions.  This naturally explains both the absence of ongoing star formation and the gas-poor state observed in YMCs \citep{Bastian2013I}. Therefore, the AGB scenario provides a plausible explanation for why YMCs do not show any evidence of MSPs.
	
    The magnetic field is another crucial factor that may affect star formation in galaxies. It pervades the Universe on various scales from the cosmic web to galaxies, stars, and planets. Early magnetic fields were probably produced due to plasma instabilities, which were then amplified by some dynamo mechanisms \citep[see e.g.,][and references therein]{subram,Beck2013,Beck2015} and gas compression \citep{Hosseini,Taba_16} to the strengths observed today in galaxies. The role of magnetic fields in the formation of star clusters has been widely discussed and recognized as an essential factor \citep{Mestel1956, Shu1987,Price2008,Price2009,Krumholz2019,Pattle2023,Dobbs2021}. 
     Magnetic fields appear to play a dual role in star formation. They can facilitate mass inflow by removing angular momentum, but can also prevent collapse by exerting the magnetic pressure \citep{Kumssa_Tessema_2018}. Observations also indicate that magnetic fields can change the pressure balance of the interstellar medium \citep{Beck_07,Taba_08,Taba_22} and, in some conditions, decelerate the formation of stars \citep{Taba_18}. Moreover, they can influence gas accretion and ejection processes in star-forming filaments, depending on the field orientation \citep{Mignon-Risse2023,Pattle2023}. 

	The role of magnetic fields in SG formation within GCs remains  unexplored. However, there are reasons to expect that magnetic fields may play an important role in this process. Previous studies have shown that magnetic fields can significantly affect the gas dynamics around moving objects, such as star clusters traversing a magnetized medium \citep{Lai2006,Lai2008,Wang2020,Mackey2007,Zhao2015, Lai2024}. In the context of SG star formation, these magnetic field effects could be particularly relevant. Moreover, the essential requirements of SG formation models, such as gas accretion and the retention of stellar winds, may be influenced by the presence of magnetic fields.
	Notably, observations of high-redshift galaxies suggest that the magnetic fields present during the epoch of globular cluster formation can be quite strong \citep{Fletcher2004,Wolfe2008,Geach2023}. This raises the possibility that magnetic fields may have played a role in the formation and properties of MSPs within these ancient stellar systems.
	Therefore, it is valuable to explore their role through magnetohydrodynamics simulations that also model other important processes, such as star formation and stellar feedback.

	In this study, we aim to investigate the hitherto unexplored role of magnetic fields on the formation and properties of SG stars within a typical massive star cluster. We first describe the simulation setup in \secref{sec:setup} and present our results in \secref{sec:results}. In \secref{sec:discussion}, we discuss our results, compare them with observations, and finally present our conclusions in \secref{sec:Conclusions}.
	\section{Simulation setup}\label{sec:setup}	
	In this paper, we study the formation of a second generation of stars within a massive star cluster, focusing on the combined effects of magnetic fields and radiation feedback, using 3D radiation-magnetohydrodynamical (RMHD) simulations.
	To achieve this we use the \ramsesrt code \citep{rosdahl2013, Rosdahl2015}, a radiation hydrodynamics extension of \ramses \citep{teyssier2002}.
	
	The setup of our simulation is very close to that in \cite{Yaghoobi2024}, hereafter \citetalias{Yaghoobi2024}, but with the difference that we include magnetic fields using the numerical implementation described in \citet{Fromang2006}. In this study, we focus on the {\sc M6Infall23} model of \citetalias{Yaghoobi2024}; we consider a cluster with the mass of $10^6\Msun$ and half-mass radius of $4\pc$ that supersonically moves through a homogeneous ISM with a gas density of $10^{-23}\gocm$ and a constant velocity of $23\kms$ in the negative x-direction. The cluster mass is comparable to initial masses derived for proto-GCs \citep{Baumgardt2019}, as here we investigate the SG formation in the first $100\Myr$ of the GC formation based on the AGB scenario. In addition to magnetic fields, our simulations include self-gravity, stellar winds, star formation, ionizing radiation, and gas cooling and heating processes.
	
    \subsection{Numerical Method}
	We now provide a detailed description of the numerical setup. To minimize boundary effects related to the presence of magnetic fields, we assume a larger simulation box than that used in \citetalias{Yaghoobi2024}. We find that a cube with a width of $256\pc$ and periodic boundaries\footnote{When gas crosses one boundary of the simulation box, it re-enters the box from the opposite side.} effectively achieves this goal. However, we consider the left and right sides (along the x direction) of the box to be inflow and outflow boundaries, respectively, to work in the reference frame of the cluster. Thus, we set the cluster to be static at the center of the box, allowing the ISM gas (the same as the gas in the initial conditions) to flow into the box from the left boundary at the assumed velocity of the cluster.
	
	Our magnetized simulations are performed in 3D Cartesian coordinates
	$(x, y, z)$ and utilize adaptive mesh refinement. We mesh our $256\pc$ box to cells with a maximum size of $\Delta x_{\rm max}= 4\pc$ ($l_{\rm min}=7$ in the \ramses code) and a minimum size $\Delta x_{\rm min}= 0.25\pc$ ($l_{\rm max} = 11$). Cells are refined at each level if the cell mass exceeds $5 \Msun$, or if the cell size becomes larger than ${\rm 1/16th}$ of the local Jeans length, which is sufficient to avoid artificial fragmentation \citep[][]{Truelove1998,Greif2011}. These refinement criteria ensure the Jeans instability is properly resolved to capture the gravitational collapse of star-forming regions up to $l_{\rm max}$, at which star formation occurs. Our analysis indicates that the adopted setup is sufficient for our results to converge with reasonable changes in resolution. 
	
	Using the \ramsesrt code, we solve the radiative transfer (RT) equations with a two-moment method and the M1 closure along with the ideal magnetohydrodynamics (MHD) equations. The code uses a second-order Godunov scheme to solve the Euler equations and a particle-mesh solver to compute the dynamical evolution of star particles. The rest of the setup of this study is the same as the one presented in \citetalias{Yaghoobi2024}.
	%
	\begin{table}
		\centering
		\caption{Main parameters for the simulations performed in this paper.}  
		\begin{tabular}{lcccccl} 
			\hline
			Run & Magnetic\ field      & Magnetic              & NoRT\ / \\
			&  strength \ [$\mu$G] & field\ direction  &  RT\\		\hline \hline
			\HD  & $0$ &  -  &\NONRT \\ 
			\BxWNoRT  & $0.5$ &  $x$ &  \NONRT \\
			\BxSNoRT & $5$  & $x$&  \NONRT \\
			\BxSSNoRT  & $50$ &  $x$ &\NONRT \\
			\hline
			\RHD      & $0$   &  - &  RT \\ 
			\BxWRT    & $0.5$ & $x$&  RT \\
			\BxSRT  & $5$   & $x$&  RT \\
			\BxSSRT & $50$  & $x$&  RT \\
			\hline
			\ByWRT  & $0.5$ & $y$  &   RT \\
			\BySRT  & $5$   & $y$  &   RT \\
			\BySSRT & $50$  & $y$  &   RT \\
			
			\hline
		\end{tabular}
		
		\label{tabl1}
	\end{table}
	%
	\subsection{Initial Conditions }\label{sec:init}
	We assume that the cluster moves in an initially uniform ISM gas with a gas density and temperature of $10^{-23}\gocm$ and $500\K$,\footnote{This is an average value between the phases of Cool H\,I (CNM) and Warm H\,I (WNM) of the ISM \citep[Table 1.3 of][]{Draine2011}.} as a representative density value for star-forming regions in galaxies at high redshifts \citep{wardlow2017}. We initiate our simulations at the point when Type II SN explosions of FG stars have ceased, and the cluster has re-entered the unperturbed ISM after traversing the region affected by these explosions. At this stage, the cluster can begin to accumulate pristine gas from the ISM. The timing of this re-entry depends on the cluster's mass, ISM density, and wind velocity, as detailed in \citet{D'Ercole2016}. Following \citetalias{Yaghoobi2024}, we estimate this re-entered time to be $33.9\Myr$ after the formation of our $10^6\Msun$ FG cluster in the assumed ISM. Hence, our simulations begins at $t = 33.9 \Myr$ and conclude at $t = 100 \Myr$ after the cluster formation, approximately coinciding with the onset of Type I SNe explosions \citep{D'Ercole2008}.
	In our previous work, we modeled the bubble formed by the SN explosions, assuming the initial conditions with a hot and diffuse gas. However, in this study, we chose to neglect the bubble, as the diffuse gas would result in higher Alfvén velocities and consequently much shorter time steps, making the computations more challenging. Instead, we assume the cluster moves through a uniform ISM.  Our primary simulations have shown that the final results are not significantly affected by this change in the initial gas conditions.
	
	Since we work in the cluster's frame, both the initial gas within the box and the incoming gas from the left boundary move in the positive x-direction with a uniform velocity of $23\kms$ (the cluster's velocity). As in our previous works, the FG stars are modeled (in terms of gravitational potential but also for AGB ejecta and ionizing radiation) using a static \cite{plummer1911} density profile:
	\begin{equation} 
		\rho_{*,\rm \ FG}(r) = \frac{3 \ M_{\rm FG}}{4\pi\, r_p^3} \left(1+\frac{r^2}{r_p^2}\right)^{-\frac{5}{2}},
		\label{plum}
	\end{equation}
	where $r$ is the distance from the center of the cluster, $r_{p}=3\pc$ is the Plummer radius, and ${M_{\rm FG}}$ is the {\rm FG} cluster mass. This modeling neglects any changes in the FG cluster during the simulation. Another study where the cluster is modeled with individual stars will be presented in a separate paper (Yaghoobi et al.). Our primary results suggest that modeling such a cluster with a static potential is a good approximation for a spherical stellar system.
	
	\subsection{Magnetic Field}
	To explore the role of magnetic fields, we conduct a parameter study of the magnetic field strength and orientation. The evolution of the magnetic field is described by the induction equation,
	\begin{equation}
		\frac{\partial \vec B}{\partial t} - \nabla\times( \vec v \times \vec B) = 0,
	\end{equation}
	where $v$ is the gas velocity and $ {\vec B}$ is the local magnetic field.
	
	Typical ISM field strengths range from a few $\mg$ for dwarf galaxies and the Milky Way \citep{Chy2003,Beck2013,Beck2015} to around $30\mg$ for star-forming regions of spiral galaxies \citep{Fletcher2004}. At higher redshifts, magnetic fields as strong as several tens of  $\mg$ have been observed \citep{Wolfe2008,Geach2023}.
	Accordingly, we consider three strengths for the magnetic field: $0.5$, $5$, and $50\mg$. We assume these fields are oriented either parallel (only along the +x direction) or perpendicular (along the +y direction) to the cluster's motion. This parameter study allows us to systematically investigate how variations in the magnetic field strength and orientation affect the SG star formation process within our cluster. 
	
	\subsection{Stellar Winds}\label{sec:SW}
	Stellar evolutionary models of AGB stars suggest that the onset of AGB ejecta within a cluster occurs about $40\Myr$ after the cluster's formation \citep{Ventura2013}. Following the model described in \cite{calura19}, we assume the injection of AGB ejecta begins at $ t_{\rm AGB} = 39 \Myr$ after the cluster formation.
	\text{The} rate of injected mass by AGB stars is modeled using a function of time $t$ as:
	\begin{equation}\label{AGB_ej}
		\dot{\rho}_{ \rm AGB}(r)=\alpha \rho_{*,\rm FG}(r),
	\end{equation}
	where $\alpha= 0.065 \, ({t}/{\rm 1 yr})^{-1.01}$ is the specific injection rate. This rate is included as a source term in the Euler mass conservation equation. We employ a passive scalar to represent the He abundance in both gas and particles. For the AGB ejecta, the He abundance changes from $Y=0.36$  at the beginning of the AGB injection to $Y = 0.32$ at the end of our simulations \citep{ventura2011}, whereas for the pristine gas, we assume it to have $Y = 0.246$. Note that we ignore changes in the specific heats $\gamma$ of the cells resulting from this varying helium fraction. Additionally, we use another passive scalar to specify the AGB stellar wind fraction in every cell to investigate the contributions of AGB ejecta (and pristine gas) to SG formation. 
	\subsection{Star Formation}\label{SF}
	An essential physical process in our model setup is star formation (SF), specifically, the process of converting gas into {\textit{stellar particles}}. To achieve this, we use the model described in \citet{Rasera2006} and introduce a sink term into the Euler conservation of mass equation, expressed as:
	\begin{equation}
		\dot{\rho}_{\rm *}= -\frac{\rho}{t_*},
	\end{equation}
  	where $\dot{\rho}_{\rm *}$ is the rate of star formation density, and $t_*$ is the SF timescale which is assumed to be $0.1 {\rm\, Gyr}$ \citep{calura19}. Our model incorporates four criteria for SF: i) a gas temperature less than $2\times10^4\K$, ii) a converging gas velocity $(\nabla\cdot \textit{v}<0)$,  iii) a local Jeans length smaller than four times the finest cell width, and iv) a density higher than $4.7\times10^{-22}\gocm$.\footnote{This threshold density is determined based on the assumed resolution and minimum particle mass \citep{Yaghoobi2022a} along with the constraint that no more than $90$ percent of the gas mass can be converted into star particles.}
	
	When all these criteria are fulfilled, star particles are formed stochastically in eligible cells. The stellar particle mass is an integer multiple of ${{m}_{*} = 0.1\Msun}$, sampled from a Poisson probability distribution as described in \citet{Rasera2006}. 
	\subsection{Radiative Transfer}\label{sec:RT}
	The \ramsesrt code models the propagation of ionizing radiation and its interplay with gas via non-equilibrium thermochemistry for hydrogen and helium. In this study, we assume three photon groups (H\,I, He\,I, and He\,II ionizing photons) with mean energies of $18.2$, $33.0$, and $61.3\ {\rm eV}$, respectively, as assumed in \citet{Yaghoobi2022b}. In particular, the evolution of ionization fractions for hydrogen and helium is tracked in every cell \citep{rosdahl2013, Rosdahl2015}. Using this code, we can study the effects of photoionization heating, cooling, and radiation pressure of the ionizing radiation emitted by FG and SG stars. We stress, however, that our previous studies demonstrate that radiation pressure does not play a role in the evolution of the gas or star formation \citep[][\citetalias{Yaghoobi2024}]{Yaghoobi2022b}, whereas photoionization heating has significant effects. 
	
	In our simulations, the FG and SG luminosities are given by the binary population and spectral synthesis (BPASS) code \citep{BPASS2017}, assuming a metallicity of $Z=0.001$ for both generations. In line with the standard approach in the study of SG star formation, we consider a truncated \citet{kroupa2001} IMF with a maximum mass of $m=8\Msun$ for the SG population, as assumed in our previous works \citep{calura19,Yaghoobi2022a, Yaghoobi2022b,Yaghoobi2024}. We, therefore, model a relevant luminosity for the SG stars with such a truncated IMF using BPASS luminosities, as described in detail in \cite{Yaghoobi2022b}. Finally, a reduced speed of light of $0.002c$ is considered to minimize the computational cost of the simulations \citep{rosdahl2013}.
	
	\subsection{Cooling and Heating}\label{sec:cooling}
	In our simulations, Euler's energy conservation equation accounts for gas heating and cooling processes.
	Radiative cooling and heating rates are functions of the gas temperature, density, photon flux, and ionization fractions, as described in \citet{rosdahl2013}. As in our previous studies, the metal cooling at lower temperatures is modeled following \citet{Rosen1995ApJ}, with a temperature floor of $100\K$. Consistent with the setup modeled by \citet{calura19}, we include the heating effects of AGB stellar winds in the energy equation as:
	\begin{equation}\label{EAGB}
		S=0.5\alpha \rho_{*,\rm \ FG}\left(3\sigma^2+v^2+v_{\rm wind}^2\right),
	\end{equation}
	where $\sigma$ is the Plummer one-dimensional velocity dispersion, which depends on the distance from the cluster's center, $v_{\rm wind}$ is the wind velocity of the AGB stars, and $v$ is the local gas velocity. We assume a wind velocity of $v_{\rm wind}=20\kms$ and consider an adiabatic index of $\gamma = 5/3$ for the ratio between internal energy and gas pressure.
    \begin{figure*}
		\centering
		\includegraphics[width=0.8\linewidth]{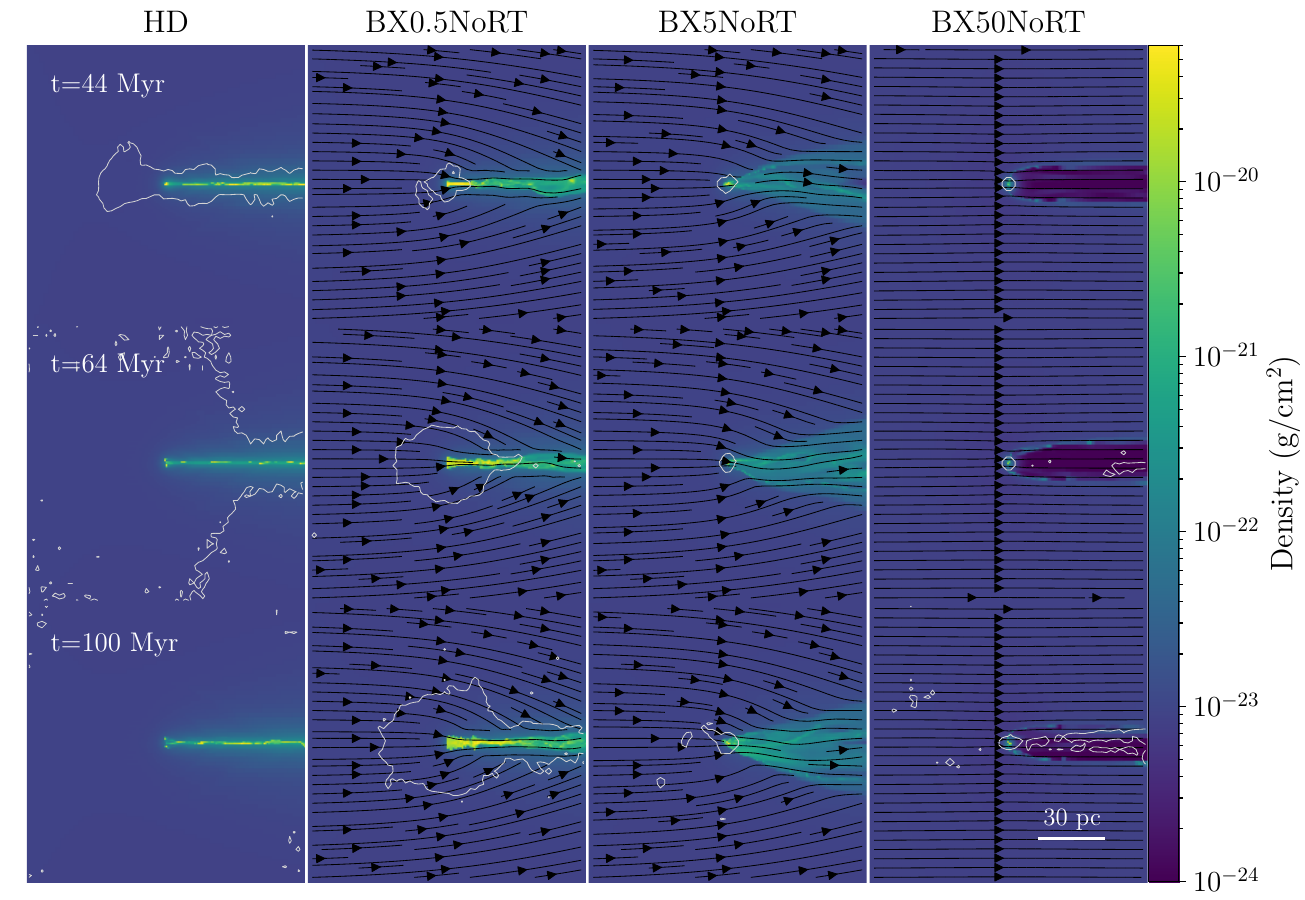}
		\caption{Gas density slices computed in the x-y plane and covered by magnetic field lines, at $t=44 \Myr$ (first row), $t=64 \Myr$ (second row), and $ t=100\Myr$ (third row) for models without photoionization feedback and with the initial magnetic field strengths of $0, 0.5, 5, 50 \mg$ (left to right) which are parallel to the inflowing gas motion. White contours indicate the extent of SG stars with a projected SG stellar density of $\sim 10^{-4}\ {\rm g\, cm^{-2}}$. We present the gas configuration for half of the box  width to emphasize the central regions. As a result, in the HD run at $t=100\Myr$, the contours cover nearly the entire displayed area. In all models, the cluster can effectively accumulate both AGB and pristine matter at its center, leading to SG star formation. However, the magnetic field significantly influences the gas morphology and star-forming regions in the shocked region behind the cluster.}
	\label{fig:NoRT}
\end{figure*}

\begin{figure*}
\centering
\includegraphics[width=0.7\linewidth]{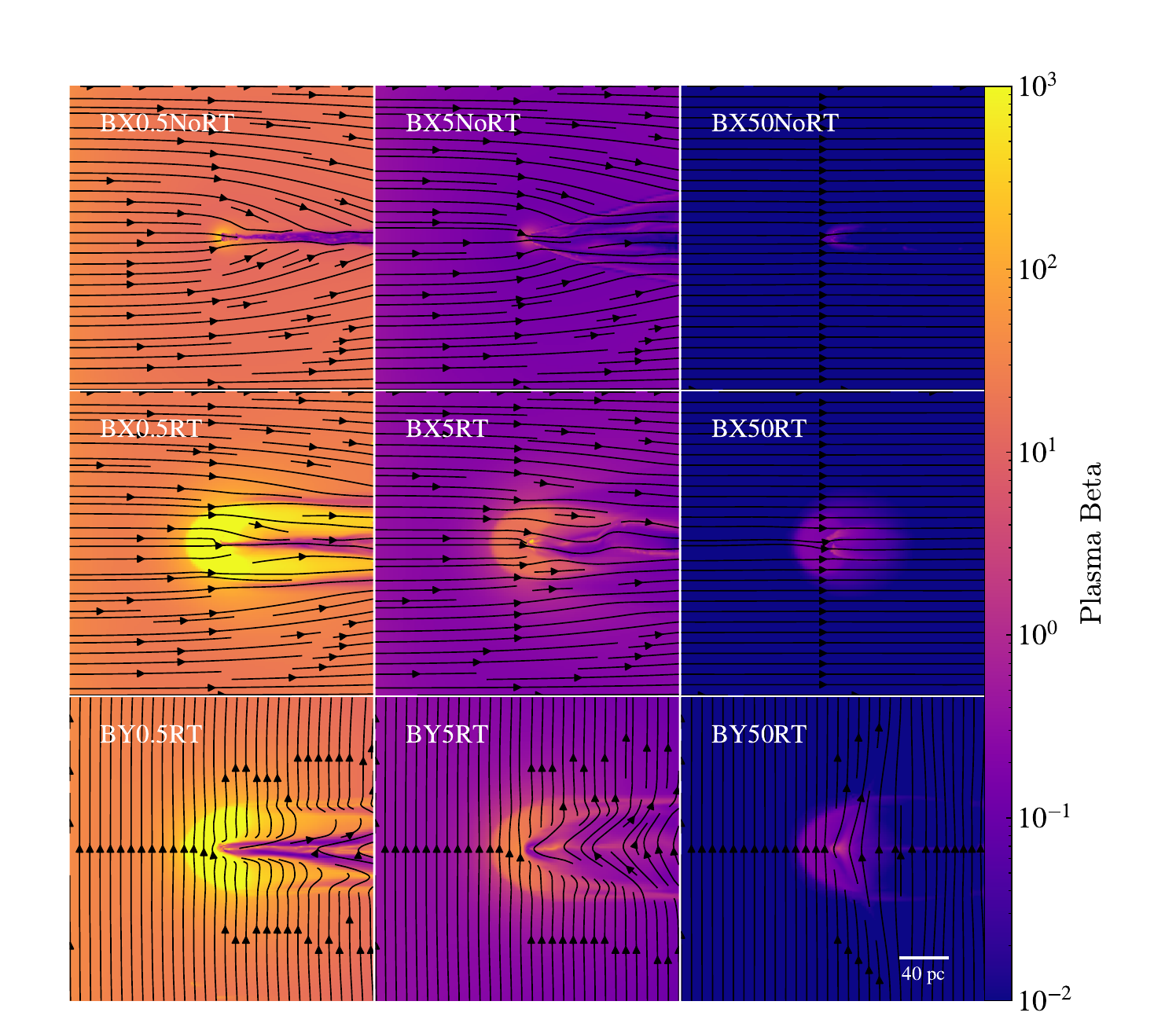}
\caption{The plasma beta slices in the x-y plane, covered by magnetic field lines, for our magnetized runs with the initial magnetic field strengths of $0.5$ (first column), $5$ (second column), and $50 \mg$ (third column) at the end of simulations ($100 \Myr$). The first row corresponds runs with magnetic fields parallel to the gas motion but without ionizing radiation. The second and third rows correspond to runs with both ionizing radiation and magnetic fields, aligned parallel and perpendicular to the cluster's movement, respectively.}	
\label{fig:beta}
\end{figure*}
\begin{figure}
	\centering
	\includegraphics[width=1\linewidth]{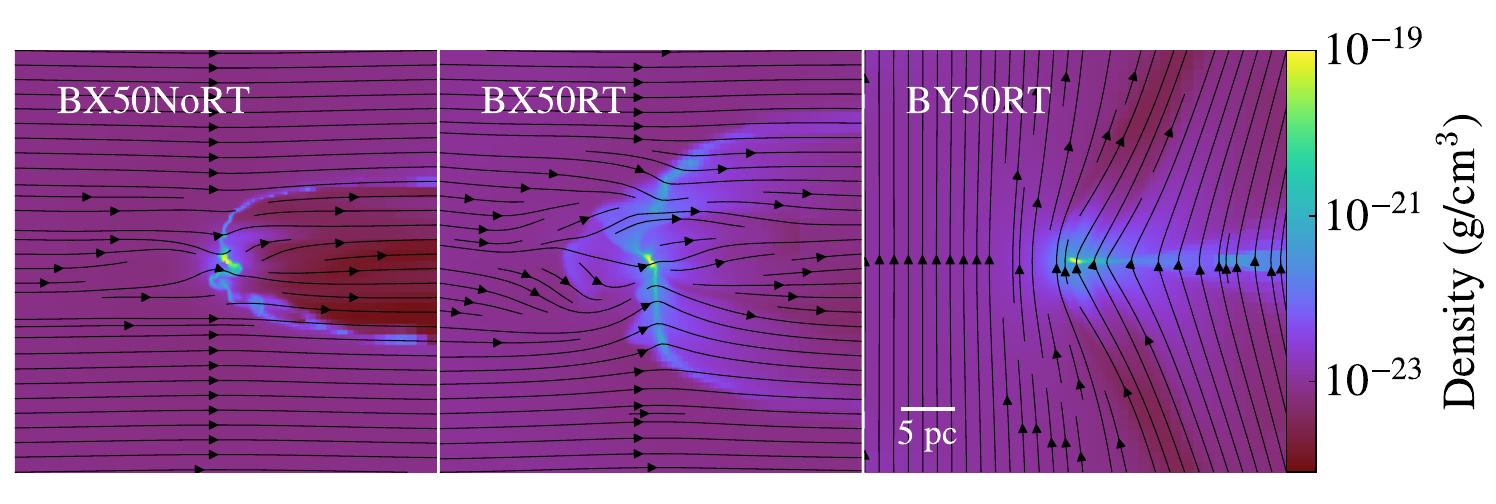} 
	\caption{Zoomed-in gas density maps on the plane $z=0$ for our strong field models. In all cases, the gas can collapse along the direction of the background magnetic field, but not perpendicular to it. Consequently, the magnetic field lines are dragged along with dense gas.}

\label{fig:zoom}
\end{figure}
\begin{figure*}
\centering

\includegraphics[width=0.8\linewidth]{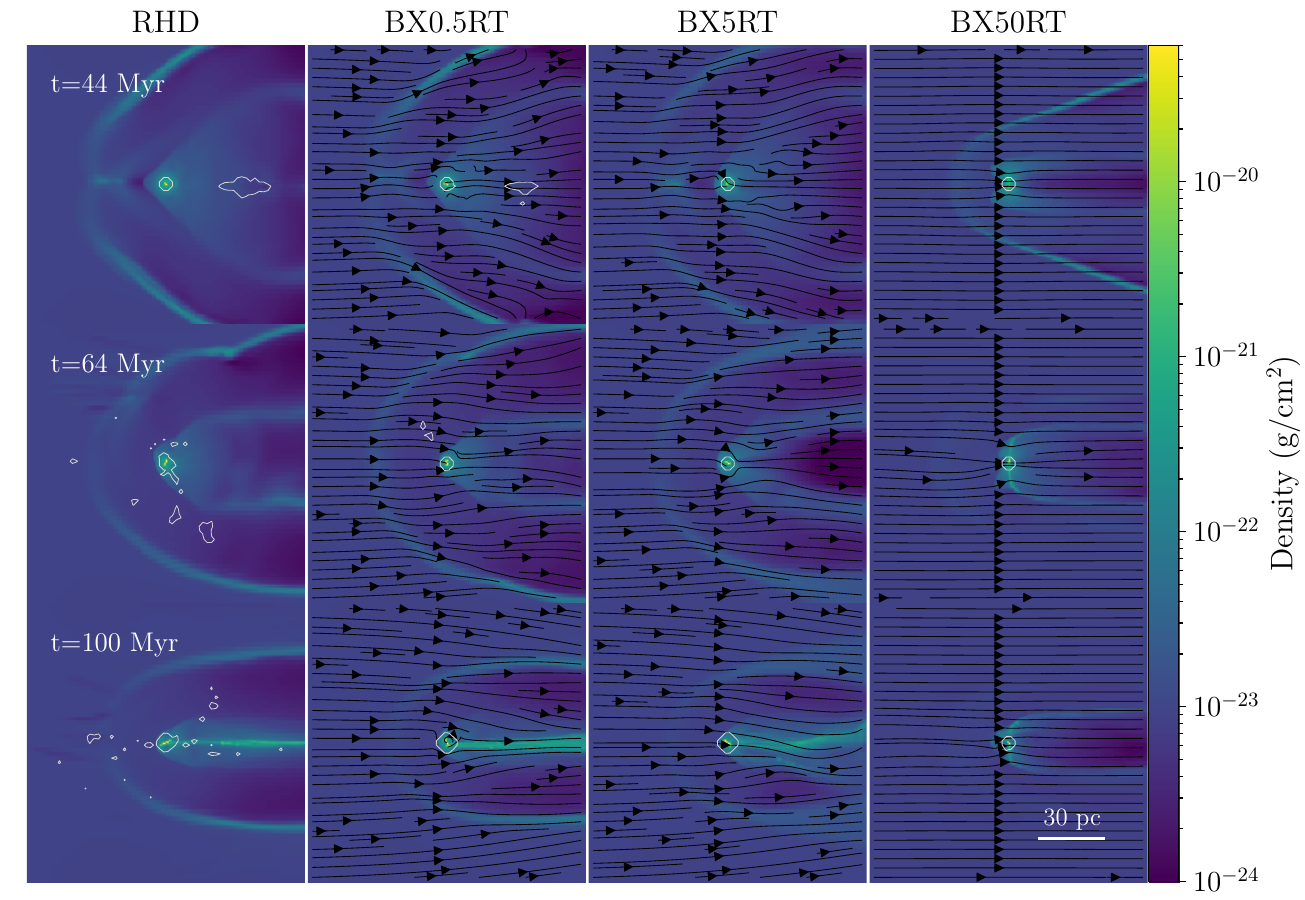}
\caption{
Gas density slices computed in the x-y plane and covered by magnetic field lines, for models including both photoionization feedback and initial magnetic field strengths of $0, 0.5, 5, 50 \mg$ (left to right) parallel to the inflowing gas motion. White contours indicate the extent of SG stars with a projected SG density of $\sim 10^{-4}\ {\rm g\, cm^{-2}}$. The photoionization radiation heats the gas. As a result, the ISM gas is pushed away, and a bow-shaped dense shell is created around the cluster. This shell specifies the region affected by ionizing radiation.}
\label{fig:BXRT}
\end{figure*}
\section{Results}\label{sec:results}
This section presents the results of our simulations, which include radiative transfer and magnetic fields. We perform some simulations without ionizing radiation to examine the influence of the magnetic field in isolation. Our simulations are summarized in \tabref{tabl1}. The first set of runs includes the magnetic field without radiative transfer (referred to as NoRT). The second and third sets incorporate both ionizing radiation and magnetic fields, with the fields aligned parallel and perpendicular to the cluster’s movement, respectively.

In our notation, time $t=0$ corresponds to the birth time of the FG stars.  Hence, we simulate the evolution of the gas and magnetic fields, as well as the star formation processes, over a time interval of $33.9$ -$100\Myr$ after cluster formation.

\subsection{Gas Dynamics and Magnetic Field Evolution}\label{sec:gas}
We begin by examining the effects of magnetic field and radiative feedback on the evolution of gas, magnetic field, and star-forming regions around the cluster. Each subsequent subsection focuses on a specific set of simulations.
\subsubsection{Magnetized simulations - no radiative feedback}
We first consider simulations that include magnetic fields without radiation feedback. \figref{fig:NoRT} displays the gas density evolution in the x-y plane for these simulations with initial magnetic field strengths of $0$, $0.5$, $5$, $50\mg$, from left to right, which are initially parallel to the inflowing gas motion. The field lines are displayed by black lines. Additionally, white contours encapsulate a projected stellar density of SG stars above $\sim  10^{-4}\ {\rm g\, cm^{-2}}$.

The leftmost column of \figref{fig:NoRT} presents the results of the hydrodynamics (HD) run, as a control simulation for the MHD runs.
In the absence of radiation and magnetic fields, the gravitational potential of the cluster is sufficient to overcome ram pressure, enabling the cluster to accrete pristine gas from the surrounding ISM. As a result, a dense gaseous tail forms behind the cluster, leading to significant star formation both at the cluster's center and along the tail, as illustrated by contours at various times. This gaseous tail is a result of the combined effects of various physical processes, such as self-gravity, cluster gravity, gas cooling, and wave propagation, acting on the gas behind the cluster.

The gas evolution for the three magnetized runs of \BxWNoRT, \BxSNoRT, and \BxSSNoRT, is presented in the other columns of \figref{fig:NoRT}, arranged in order of increasing magnetic field strength from left to right. 
The panels in the second column illustrate the gas evolution in the \BxWNoRT run. In this case, the tail forms as the gas passes through the center of the cluster. A weak magnetic field aligned with the gas motion results in a tail that extends in the y-direction compared to the simulation without magnetic field. This extension can be attributed to the increased magnetic pressure and the propagation of fast magnetosonic waves perpendicular to the magnetic field within the shocked region (we will discuss this aspect further in \secref{sec:discussion}). Consequently, the star formation region at $t=44\Myr$ is limited to the center of the cluster, in contrast to the HD run, and subsequently expands into the outer regions of the tail at later times.
The magnetic field lines align with the gas motion, originating from the left along the x-direction, tilting toward the tail, and continuing horizontally within it.

To evaluate the impact of the magnetic field relative to gas pressure, we calculate the plasma beta parameter  $\beta$, which represents the ratio of thermal pressure to magnetic pressure. In \figref{fig:beta}, we present this parameter in the x-y plane for our magnetized simulations at the final time ($t=100\Myr$), along with overlaid magnetic field lines. 
Considering the figure, it can be seen that in the run with a weak magnetic field (top-left panel), the thermal pressure dominates across most regions ($\beta>10$), except within the gas tail. In contrast, outside the tail, the magnetic field remains too weak to have a significant impact on the gas dynamics.

In the \BxSNoRT case, the cluster moves (at a velocity of $20 \kms$) through the ISM, which has a higher Alfvén speed of $4.5 \kms$. As a result, fast magnetosonic waves propagate more quickly in the y-direction than in the $0.5\mg$ magnetic field case. This leads to the formation of a cone-shaped shock with a larger angle behind the cluster, effectively separating the regions that have been shocked by the cluster's movement from the non-shocked background. In addition to the gas, the magnetic field lines also expand within the shocked regions, resulting in smoother amplifications in magnetic field configurations compared to those observed in the weak-field case. The center of the FG cluster is characterized by a dense, cold gas, with a density of approximately $\approx 10^{-19}\gocm$ and a temperature of $100 \mathrm{K}$. These conditions are conducive to the formation of SG stars. Additionally, dense filaments aligned with the magnetic field are formed within the shocked region, leading to the formation of some SG stars within that area.
\begin{figure*}
\centering
\includegraphics[width=0.9\linewidth]{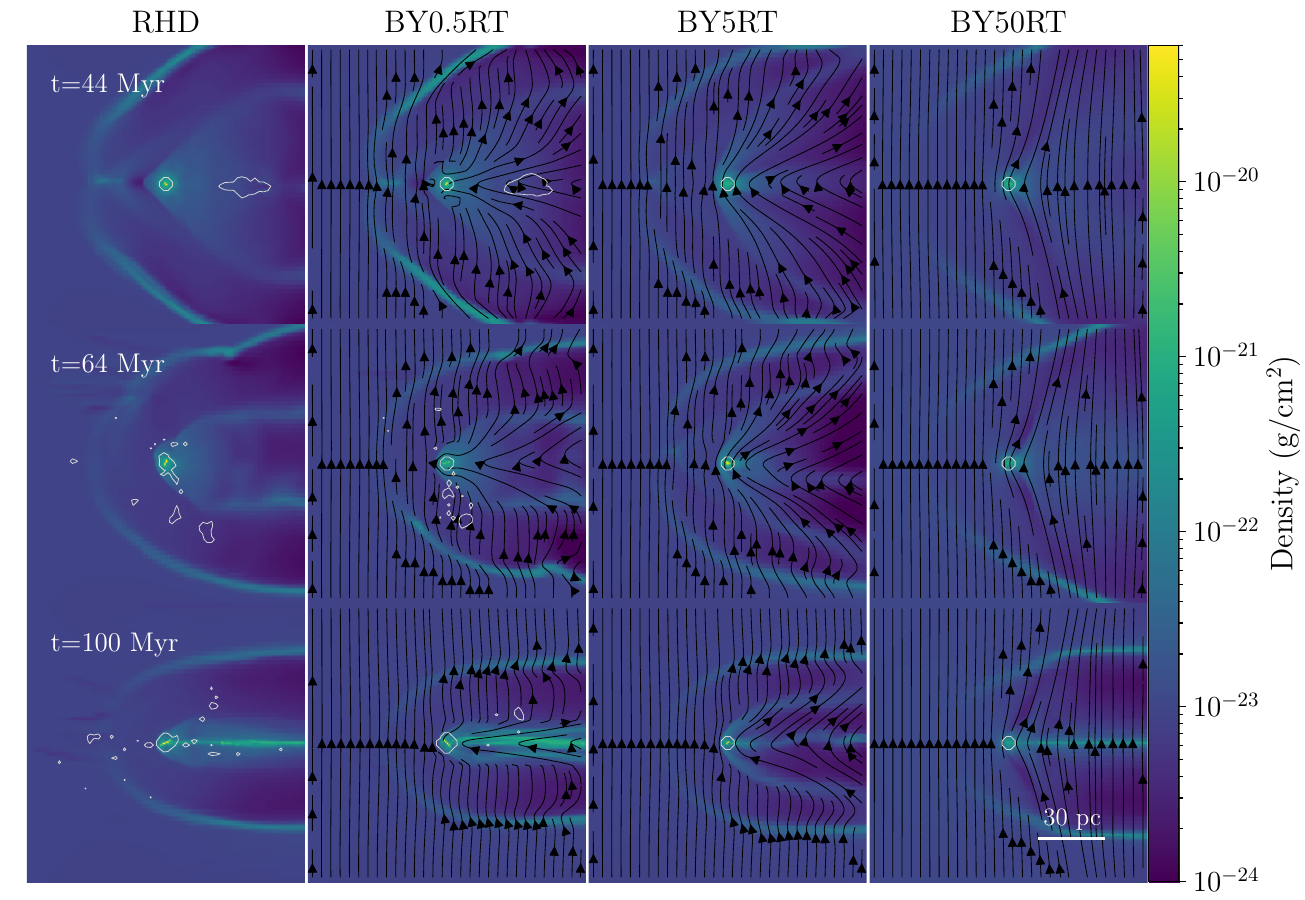}
\caption{Similar to Figure \ref{fig:BXRT}, but this shows cases where the magnetic fields are oriented perpendicular to the inflowing gas motion. In this configuration, varying the magnetic field orientation (along the y-axis) results in distinct evolutions of gas morphology and magnetic field structures compared to those observed with parallel fields.}	
\label{fig:BYRT}
\end{figure*}

In the extreme case (\BxSSNoRT) where the magnetic field strength completely dominates in the box, the beta parameter falls below unity, as shown in the top right panel of \figref{fig:beta}. 
The fourth column of \figref{fig:NoRT} illustrates how the magnetic field lines remain unchanged inside the box and the magnetic field governs the gas dynamics.
Consequently, only the gas passing through the central regions can respond to the cluster's potential. In other regions, the strong magnetic field prevents the gas from moving vertically, due to the strong coupling of the plasma to the field lines.  As a result, the gas accretion into the cluster is limited to the very central regions where the gravitational pull of the cluster is very strong.
The central gas can collapse horizontally, but not perpendicular to the magnetic field, as shown in the left panel of \figref{fig:zoom}.  The magnetic field lines are also dragged along with the dense regions. These conditions still provide a favorable environment for star formation at the center of the cluster, although the process is more confined to a small area. Behind the cluster, higher magnetic pressure leads to the formation of a low-density tunnel surrounded by a higher-density thin shell (right panels of \figref{fig:NoRT}). As a result, some new stars are formed in this region, which is why the SG contours appear elongated behind the cluster.  
\subsubsection{RMHD simulations ($B_x$ cases)}\label{sec:RMHDx}
In this section, we examine the evolution of the gas while accounting for both stellar radiation feedback and the influence of the magnetic fields parallel to the gas motion.	\figref{fig:BXRT} shows the gas evolution from these RMHD simulations. Compared to the HD case, the radiative hydrodynamic (RHD) run exhibits two distinct gaseous structures: a warm, lower-density region and an inner shock within it. The photoionization radiation heats the cluster gas, causing the gas pressure and temperature to increase. As a result, the ISM gas is pushed away, forming a dense, bow-shaped shell around the cluster. This shell confines the region affected by the ionizing radiation. At the beginning of the simulation, the shell forms spherically. As the gas flows through the computational domain from the left side, the shell shifts towards the right. Near the cluster, the tip of the shell becomes unstable; this instability causes the shell to deform, leading to a gas accumulation at its apex. The accumulated gas then moves inward toward the cluster’s center, as illustrated in the top-left panel of Figure \ref{fig:BXRT}. The inner bow shock is created by the cluster's motion relative to the ionized gas. Consequently, due to the photoionization feedback, the cluster moves through a medium with different properties ($\rho \approx10^{-24}\gocm$ and $T\sim 10^4\K$) from those of the background. We provide temperature panels for all runs at the end of each simulation in Figure \ref{fig:temp} of Appendix \ref{Appendix1}.

As the simulation progresses, the luminosity of the cluster decreases \citep{Yaghoobi2022b}, causing both the outer and inner shocks to shrink, which results in the formation of the tail at  $t=100 \Myr$. However, this tail is not as narrow as the one observed in the \NONRT run. Consequently, the radiative heating, which alters the temperature and gas density, plays a significant role in the gas morphology of the shocked regions around the cluster and leads to a temporary delay in tail formation. While the radiation does not suppress the accumulation of gas at the center, it does affect the tail until approximately  $t=80 \Myr$. As a result, star formation begins at the center of the cluster, followed by the formation of some stars in the shocked regions. We consider this model as a control run for the RMHD simulations.

The \BxWRT model shows that the overall morphology of the shocked regions is similar to the RHD case. However, we do observe a slight dissipation of the gas in the shell and tail at the end (see \figref{fig:BXRT}, second column). The final tail also exhibits a slightly larger opening angle than RHD, due to the propagation of fast magnetosonic waves in the y-direction. 
On the other hand, adding radiation to the \BxWNoRT case causes the gas to expand due to radiative heating, resulting in an increase in plasma beta to values greater than $500$ in front of the cluster and $0.7$ behind it (see the middle-left panel of \figref{fig:beta}). This indicates that the gas energy dominates the magnetic energy within the cluster. Consequently, compared to simulations without radiation feedback, the magnetic field becomes less effective in influencing gas dynamics, particularly in ionized regions. However, as the simulation proceeds and the radiative effects decrease, the magnetic fields amplify further. The contours show that the SG formation generally occurs in the central region with a few stars along the final tail. 

The panels in the third column of \figref{fig:BXRT} show the effects of a $5\mg$ magnetic field on the gas morphology within the cluster. This stronger magnetic field causes the shell and inner shock to diffuse less in the y-direction, likely due to the increased magnetic pressure.
Furthermore, the middle panel of \figref{fig:beta} indicates that the plasma beta parameter decreases to around $10$ upstream of the cluster and $\sim 0.2$ downstream. Again, this suggests that the magnetic field is dominating the plasma dynamics in the shocked regions behind the cluster. However, in the central regions where the SG stars form, the gas pressure almost matches the magnetic pressure, as observed in the weak-field case. These variations in the plasma beta across the system result in a more limited tail and a blurred shell structure behind the cluster.

In the \BxSSRT run, the bow-shaped shell appears at the beginning of the simulation due to stellar feedback and the motion of the cluster. Then, it moves toward the right along with the inflowing gas and leaves the simulation box at approximately $t=52\Myr$ (see the top right panel of \figref{fig:BXRT}). After that, the global structure of the gas and magnetic field lines is identical to the simulation without radiation. The magnetic field predominantly governs the gas dynamics, with only the gas passing through the central regions responding significantly to the cluster’s potential. However, thanks to ionizing radiation, the effectiveness of the magnetic field is somewhat reduced compared to the case without radiation. As seen in middle-right panel of \figref{fig:beta}, in the presence of photoionizing radiation, the plasma beta increases to approximately $0.3$ in the central regions of the cluster up to a radius of $\sim 50\pc$. This allows the central gas over a larger region than in the BX50NoRT run to collapse horizontally, leading to a more pronounced expansion of the dense central region along the y-axis. Consequently, this results in an increase in gas accretion in this model compared to the other simulations. Therefore, including a strong magnetic field parallel to the cluster's motion in the RHD run somewhat enhances gas accretion.

\subsubsection{RMHD simulations ($B_y$ cases)}
In this section, we explore how a different magnetic field orientation influences the evolution of both gas and magnetic fields. In \figref{fig:BYRT}, we display the gas evolution of simulations including the radiation feedback and magnetic field perpendicular to the gas motion ($B_y$). 

The panels in the second column of \figref{fig:BYRT} show the gas evolution for the \ByWRT model at the same time intervals as the previous runs. The general morphologies of the outer and inner shocks at $44$ and $64$ Myr are similar to those observed in the \BxWRT model. However, the magnetic field morphology differs significantly. In the \ByWRT run, the magnetic field is further amplified by the gas dynamics in the shocked region, resulting in field strengths exceeding $10\mg$ behind the cluster by the end of the run. This difference in magnetic field evolution leads to stronger magnetic tension and pressure forces in the \ByWRT simulation. The enhanced magnetic effects result in a more extended tail behind the cluster compared to the \RHD and \BxWRT models. The plasma beta figure further demonstrates this effect, showing that the vertical magnetic field is more effective in diffusing the gas in the y-direction within the shocked region.
\begin{figure*}
\centering
\includegraphics[width=0.9\linewidth]{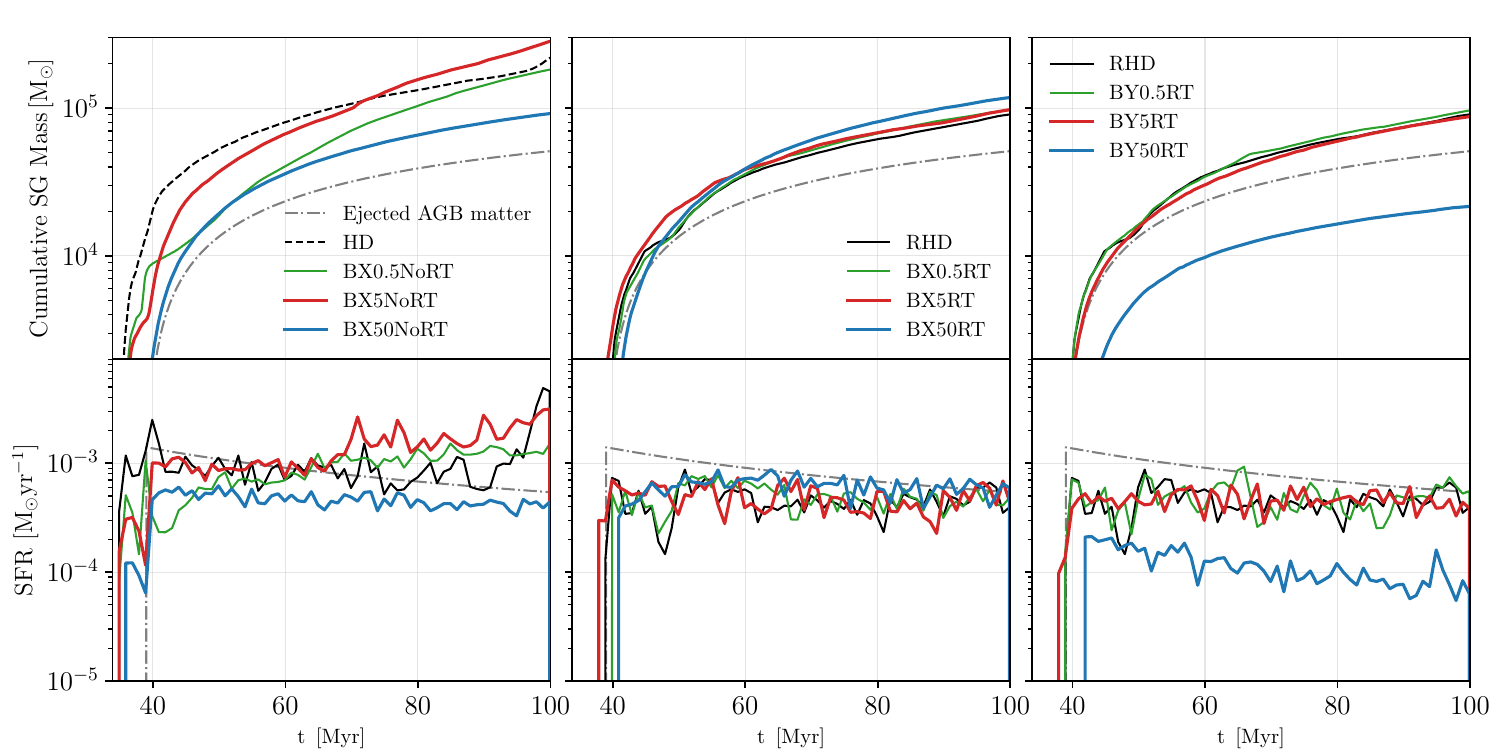}
\caption{Top panels: cumulative masses of SG stars formed in the \NONRT (left), parallel \RT magnetized ($B_x$) (middle), and perpendicular \RT magnetized ($B_y$) simulations as a function of time. Bottom: SFR of the SG stars versus time for these runs. The dash-dotted lines in both the top and bottom panels represent the cumulative injected mass and the corresponding injection rate into the simulation box, respectively.}
\label{fig:SFR}
\end{figure*}
\begin{figure*}
\centering
\includegraphics[width=0.9\linewidth]{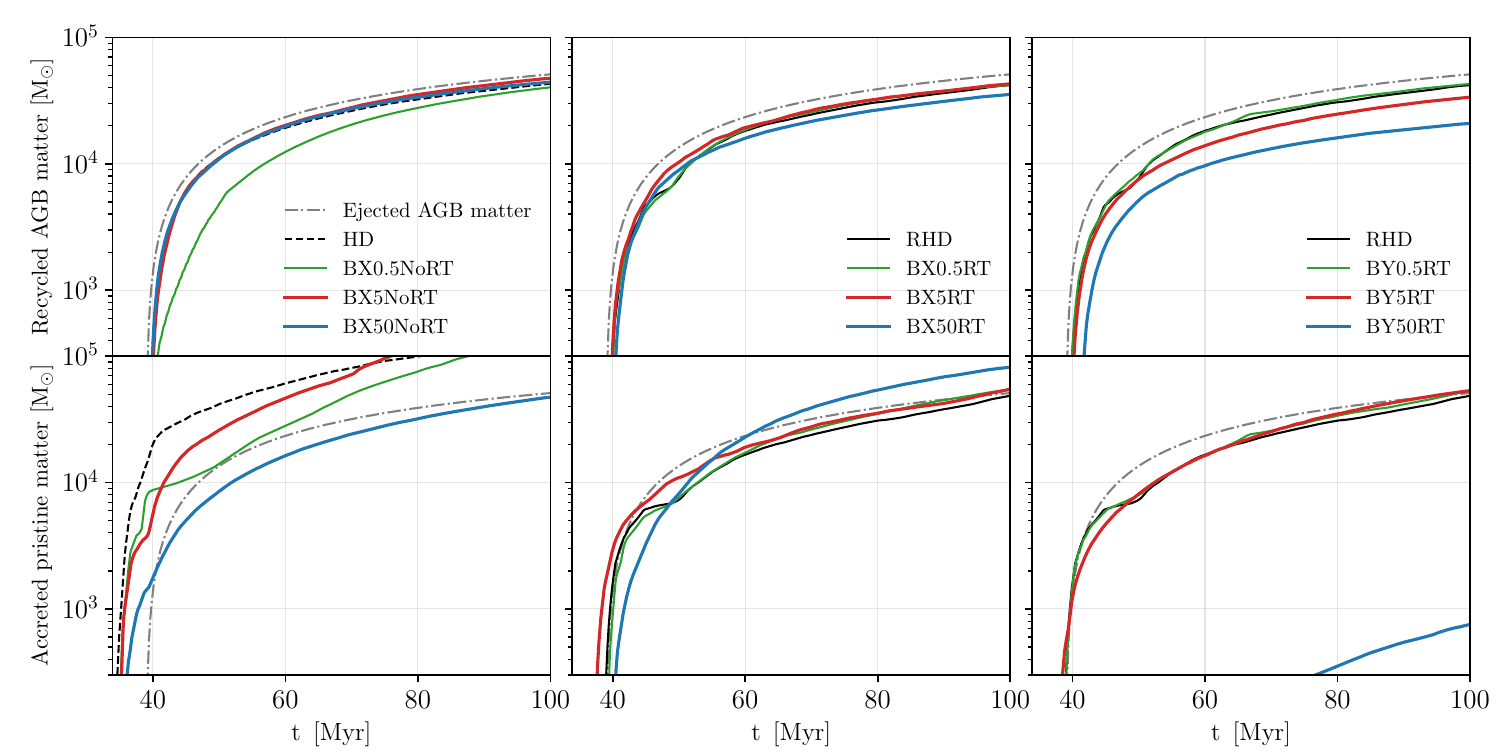}
\caption{Cumulative AGB ejecta mass converted to SG stars (top panels) and same for accreted pristine gas (bottom panels) in the NoRT (left), RHD with parallel magnetic fields (middle), and RHD with perpendicular magnetic fields (right) runs, as a function of time. The total AGB matter injected into the simulation box is also plotted for comparison.}
\label{fig:dilution}
\end{figure*}

The \BySRT case exhibits some key differences compared to the \BxSRT scenario. The higher magnetic energy and stronger coupling between the gas and magnetic field results in more gas dissipation along the y-direction. This causes the inner shocked region to elongate further along the magnetic field lines, 
forming a lower-density area behind the cluster. Additionally, the higher magnetic pressure gradient and vertical magnetic tension force at the inner shock position lead to a larger opening angle for the inner shock. This contributes to the absence of a dense tail behind the cluster in this case. However, the magnetic tension force at the axis of symmetry pushes gas in the +x direction, resulting in a thin gaseous tail, see the magnetic field lines in Figures \ref{fig:zoom} and \ref{fig:beta}. Nevertheless, the SG star-forming regions are smaller than those with the parallel $5\mg$ magnetic field.
It should be noted that the shape of the shocks arises from the interplay of various processes, including gas motion, gravity, magnetic pressure gradients, Lorentz forces, and ionizing radiation. The plasma beta parameter analysis shows that in certain central regions of the \BySRT case, the beta is less than $0.5$, indicating the magnetic field is more dominant compared to the \BxSRT scenario.

The \BySSRT simulation reveals markedly different gas structures (see the relevant panels in \figref{fig:BYRT}) compared to the similar run with a parallel field orientation.  Throughout the simulation, the ionized regions behind the cluster remain separated by a denser layer than the background (also see \figref{fig:temp}). As seen in \BySRT, at the symmetry axis a more pronounced tail by the end of the simulation is formed due to the tension force along the x direction. However, the central gas exhibits lower density relative to the other simulations, likely due to the intensified effects of this strong magnetic field. Looking at the magnetic fields lines in the x-y plane (the bottom left panel of \figref{fig:beta}), the magnetic field behind the cluster experiences greater amplification compared to the \BxSSRT case, though this amplification is still mild compared to runs with weaker or moderately stronger perpendicular magnetic fields.

\begin{figure*}
\centering
\includegraphics[width=\linewidth]{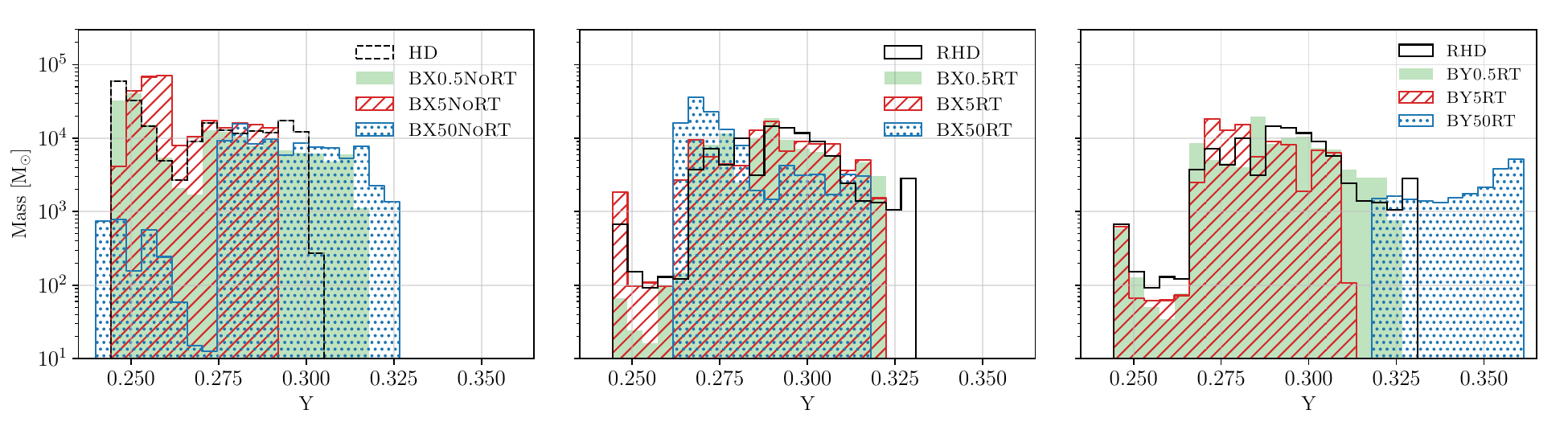}
\caption{Mass distribution of SG stars at the end of simulations versus He abundance ${\rm Y}$ for our simulation. The He abundance of the SG stars varies between ${\rm Y} = 0.24$, corresponding to the SG stars formed from the pure pristine gas, and ${\rm Y} = 0.36$, corresponding to the SG stars formed from the AGB ejecta.}
\label{fig:distrib}
\end{figure*} 
\subsection{SG Star Formation}\label{sec:SFR}
The top and bottom panels of \figref{fig:SFR} present the cumulative SG mass and star formation rate (SFR) over time for our three series of magnetized simulations, as detailed in \tabref{tabl1}. 

The magnetized simulations are represented by different colors in each column of the figure. Additionally, the dash-dotted lines in the top and bottom panels indicate the cumulative injected mass and the injection rate of the AGB ejecta, respectively.

As shown in the top-left panel of the figure, a new stellar generation with a mass of $2.2\times 10^5\Msun$ is rapidly formed in the pure HD simulation (black dashed line) in the absence of opposing forces from radiation pressure or magnetic fields.  Adding parallel magnetic fields with the strengths of $0.5$, $5$, and $50\mg$ results in the SG masses of $1.7$, $2.7$, and $0.9\times 10^5\Msun$, respectively. This indicates a non-linear behavior in SG mass for different magnetic field strengths, arising from the complex interplay of gas, gravitational force, ram pressure, and magnetic fields. We highlight that the most massive SG cluster is found in BX5NoRT. Additionally, SG stars are significantly more concentrated at the center of the cluster compared to HD and BX0.5NoRT (\figref{fig:NoRT}). This suggests that the SG density in this run is extremely high.
In general, magnetic fields have fairly strong effects on star formation in the absence of radiation feedback. The bottom-left panel of \figref{fig:SFR} shows that the SFRs exhibit different trends for NoRT runs. The first SG stars form only from pristine gas, with a He abundance of $0.246$. After the beginning of  AGB ejecta injection at $t = 39 \Myr$, the SFRs quickly increase and then tend toward a nearly constant rate with different fluctuations.

The middle panels of \figref{fig:SFR} show the development of SG stellar mass under conditions with radiation and magnetic fields oriented along the x-axis.
The black line represents the pure RHD simulation. Hence, adding radiative stellar feedback to the HD run (black dashed line) results in an approximately $5 \Myr$ delay in star formation, with a final SG mass of $0.92\times 10^5\Msun$. This represents about a $50$ percent decrease in the total SG stellar mass due to radiative feedback effects, as computed in \citetalias{Yaghoobi2024}. This is a consequence of the reduced amount of accreted material.

Incorporating magnetic fields into the RHD simulation has a different impact on the final SG masses compared to radiative feedback. For magnetic field strengths of $0.5$, $5$, and $50\mg$, the final SG masses are approximately $1.0$, $1.0$, and $1.2 \times 10^5\Msun$, respectively. Compared to the pure RHD case, there is an $8\%$ increase in SG mass for the \BxWRT and \BxSRT run and a $26\%$ increase for BX50RT. The larger increase in SG mass in the \BxSSRT run is due to the formation of a extended dense area along y-direction, resulting in a larger cross-section for gas accretion, as discussed in Section \ref{sec:RMHDx}. The bottom middle panel of \figref{fig:SFR} shows that the SFR variations seen in the \NONRT runs are diminished, and all runs exhibit similar SFR evolution trends.

In the RMHD simulations with magnetic fields perpendicular to the cluster's motion, the overall trends for the SG mass evolution and SFR remain nearly identical for the $0.5$ and $5 \mg$ field cases. The total SG mass is $0.96$ and $0.87 \times 10^5\Msun$, respectively, for these weaker field strengths (top right panel of \figref{fig:SFR}). However, a significant decrease in SG mass is observed in the $50\mg$ field run, with a final SG mass of only $0.21\times 10^5\Msun$. In this model, the SFR shows a delay of a few million years and a decreasing trend, in contrast to other models where it remains nearly constant. 
\begin{figure*}
\centering
\includegraphics[width=1.\linewidth]{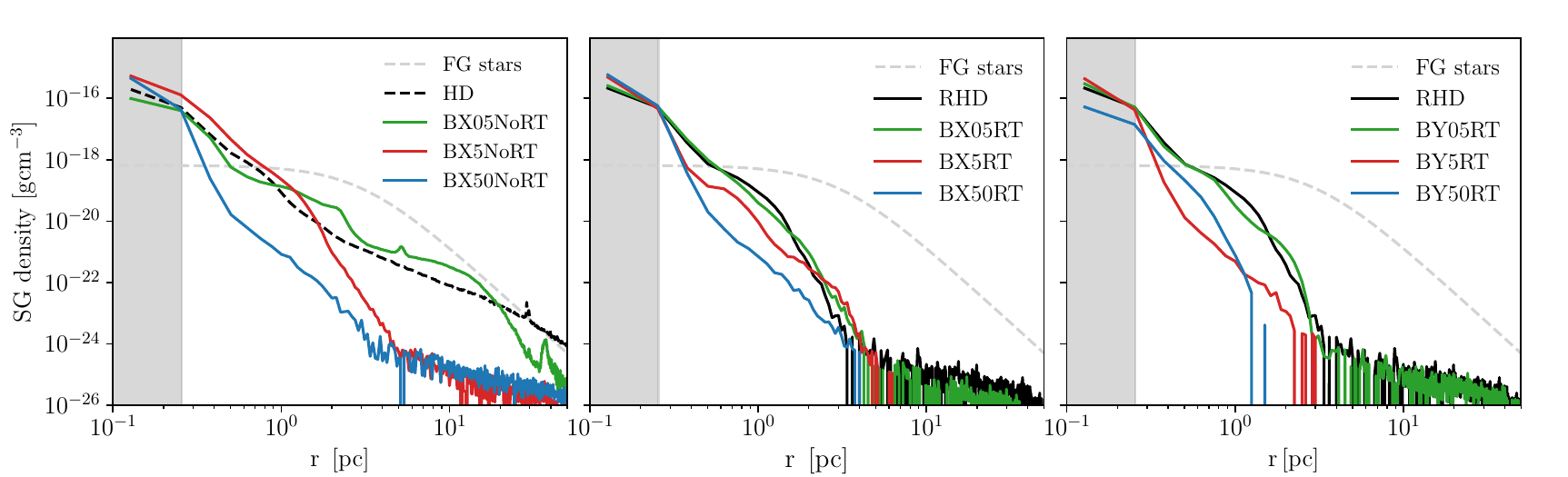}
\caption{SG stellar density profiles for NoRT (left), radiative simulations with magnetic fields parallel (middle), and perpendicular (right) to the cluster's motion. The gray dashed line represents the assumed FG profile, and the shaded area indicates the resolution of our simulation at the center.}	
\label{fig:SGprof}
\end{figure*} 

\subsection{AGB Ejecta Dilution and Stellar He Variations}\label{sec:Dilution}
To explore the role of the dilution process of AGB ejecta in SG formation during our simulations, we track the AGB mass fraction in each cell with a passive scalar. This fraction is recorded in each SG particle that forms from the gas within the cell. Our analysis indicates that, in all our RMHD simulations, stars with a high ejecta fraction and higher helium abundances predominantly form at the cluster's center. As the distance from the center increases, we observe SG stars with lower ejecta fractions and reduced helium abundances, in line with our previous studies \citep{calura19, Yaghoobi2022a}. This trend arises because the rate of AGB ejecta injection decreases with increasing distance from the cluster's center (\secref{sec:SW}) and the fact that this rate is higher than the pristine gas accretion rate at the center. In the upper and lower panels of \figref{fig:dilution}, we illustrate the evolution of the mass contributions from stellar winds and pristine matter to star formation, respectively.

In all magnetized runs without radiation (top-left panel of \figref{fig:dilution}), the cluster retains most of its ejecta and can convert them into star particles, resulting in similar contributions of AGB matter to the final SG masses. In contrast, the amount of accreted pristine matter contributing to SG formation varies with magnetic field strength, exhibiting a non-linear behavior in SG mass for different magnetic field strengths. \figref{fig:distrib} illustrates the He distribution of all SG stars formed in our simulations. The left panel, corresponding to the magnetized runs without radiation, demonstrates another non-linear behavior with increasing magnetic field strength. The less widespread distribution of Helium mass fraction in the \BxSNoRT run is attributed to more accreted pristine gas that results in a greater dilution of the enriched ejecta and, therefore, the formation of SG stars with lower He abundances.

The middle panels of \figref{fig:dilution} show that including ionizing radiation in the HD run decreases the amount of gas accretion by about a factor of 5. This reduction in gas accretion leads to a lower level of dilution for the AGB ejecta in the RHD simulation, resulting in the formation of SG stars with higher helium abundances, as illustrated by the black solid line in \figref{fig:distrib}. For magnetized runs, the amount of gas accretion at the center shows only small differences between the weak and intermediate magnetic fields. However, considerable changes are seen in the stronger magnetic field case. Specifically, a $16$ percent decrease in the retained AGB matter is observed, coupled with a $60$ percent increase in the accreted pristine material, compared to the RHD simulation, as seen in the panels of the second column of \figref{fig:dilution}.  As a result of more gas dilution, the middle panel of \figref{fig:distrib} indicates that stronger fields lead to a slightly lower maximum He abundance in the parallel RMHD simulations. 

The right panels of \figref{fig:dilution} indicate that stronger vertical magnetic fields lead to a lower contribution of stellar winds to SG mass. In contrast,  the contribution of pristine material is similar for the $0.5$ and $5 \mg$ simulations, while it is negligible in the $50\mg$ case until $t=100\Myr$. 
This discrepancy occurs because the strong field perpendicular to the cluster motion disperses the incoming gas behind the cluster more effectively in the y-direction than in the other models, as observed in the panels related to BY50RT in \figref{fig:BYRT}. Consequently, SG stars form mostly from the central AGB ejecta. Therefore, a strong magnetic field perpendicular to the cluster motion can significantly reduce the accumulation of pristine gas at the cluster center, leading to the formation of helium-rich SG stars (left panel of \figref{fig:distrib}). The maximum helium abundance reaches approximately $0.36$  in this strong field case, compared to $0.325$ and $0.31$ for the weaker and moderate field cases, respectively. 
\subsection{SG Profiles}\label{sec:SGcluster}
In order to investigate the role of magnetic fields on the distribution of SG stars formed within a massive FG cluster, we calculate the final SG density profiles for all simulations, as presented in \figref{fig:SGprof}. The FG density profile is included as a gray line, enabling direct comparison between SG and FG profiles.

For all simulations, the SG density profile is higher than the FG profile in the very central regions. Moreover, stronger magnetic fields generally result in more concentrated SG star distributions in the central regions and suppress SG formation in outer radii from the cluster's center. This leads to steeper density declines in the radial profiles. This outcome is due to the influence of magnetic fields on SG star formation regions. In environments with stronger magnetic fields, the formation of a gaseous tail is suppressed, confining SG formation to the central dense regions of the cluster. Notably, the SG profile in the \BxSNoRT case exceeds that of the weak-field run, reflecting the non-linear behavior previously noted. This model leads to the formation of the most compact SG cluster among all our models.
\section{Discussion}\label{sec:discussion}	In our previous studies \citep[][\citetalias{Yaghoobi2024}]{Yaghoobi2022b}, we investigated the effects of ionizing feedback on the formation of a second stellar generation in clusters of different masses within the context of the AGB scenario \citep{D'Ercole2008}. We considered the contributions to star formation of stellar winds from AGB stars and gas accretion from the ISM into the cluster. Our findings indicated that ionizing radiation suppresses SG formation within lower-mass clusters and lower-density environments. This is in line with the expectation that multiple generations of globular clusters have been formed at higher redshifts, when gas densities were higher than in the present-day universe \citep{D'Ercole2016}. Additionally, we found that a new generation of stars with a mass of about $10^5 \Msun$ (SG-to-FG ratio of $0.1$) can be formed within a $10^6\Msun$ cluster moving through the ISM.

In this study, we add the inclusion of magnetic fields with different strengths and orientations to investigate their role in the formation of a new generation within our $10^6 \Msun$ cluster moving through a high-density medium. Since magnetic field strengths at high redshifts remain poorly constrained, we conduct a parameter study on field strengths of $0.5$, $5$, and $50\mg$, considering two orientations: one aligned with the cluster's motion and the other perpendicular to it. This range of magnetic field strengths is proposed for lower and higher-redshift galaxies \citep{Chy2003,Beck2013,Beck2015,Wolfe2008,Geach2023}. Our simulations, incorporating both radiation and magnetic fields, reveal that the magnetic field significantly influences the gas configuration and star-forming regions around the cluster through the propagation of fast magnetosonic waves and magnetic force terms: magnetic pressure and tension. 

Nevertheless, only insignificant changes in the SG mass evolution and SG fraction are found for a magnetic field strength of $0.5 \mg$, regardless of the field orientation. For moderate magnetic fields, we find slightly different dilution,  resulting in maximum helium abundances of $0.32$ and $0.315$ for SG stars with field orientations parallel and perpendicular to the cluster's motion, respectively. These values align with observed measurements, which are below $0.315$ for such a massive cluster \citep{milone2020}.

For the case of a $50\mg$ field, we observe the strongest effects of the magnetic field on SG formation. When the field is aligned parallel to the inflowing gas, there is a $25$ percent increase in SG mass, resulting in an SG-to-FG ratio that rises from $0.1$ (for RHD) to  $0.12$. We attribute this increase to the combined effects of strong magnetic field and ionizing radiation on the central gas, which enhances the cross-section for gas accretion and promotes the formation of a denser region perpendicular to the inflowing gas.  Conversely, when the magnetic field is oriented perpendicularly, we observe a $70$ percent decrease in SG mass, leading to an SG-to-FG mass ratio of $0.02$. For
this run, almost all SG mass originates from AGB ejecta and has the maximum He abundance of approximately $0.36$. This suggests that the perpendicular orientation of the strong magnetic field significantly suppresses gas accretion onto the cluster.

Our simulations terminate at $t = 100\Myr$ after the initial cluster formation. To make a meaningful comparison between the computed SG fractions (the SG-to-total cluster mass ratio) and the observed values in globular clusters, we need to consider the long-term dynamical evolution of the cluster over several gigayears, as done in previous studies \citep{D'Ercole2008,Vesperini2021}. Due to this long-term dynamical evolution, GCs can lose a large fraction of their mass, predominantly located in the outer regions of the cluster \citep{Sollima2024,Lacchin2024,Ishchenko2025}. If the SG stars are mostly concentrated
at the cluster center, they are supposedly less affected by this mass loss \citep{D'Ercole2008, Vesperini2021, Sollima2024}. As a result, the SG-to-FG ratio can increase by a factor of around 10 over the course of the cluster's evolution \citep{renzini2015}. Considering these dynamical effects, the SG fractions of all our simulations (except for BY50RT) can increase up to $\sim 0.5$, which is in agreement with the observed SG fractions of $\sim 0.5$ to $\sim 0.8$ in such massive GCs \citep{milone2017,Lagioia2019,milone2020}. However, the BY50RT run, with the strong $50\mg$ magnetic field perpendicular to the cluster's motion, is the only exception. In this case, even after correction for dynamical evolution, the final SG fraction is only $0.16$, which is significantly lower than the observed range. This suggests that a perpendicular B-field run might very strongly affect the  SG formation. Moreover, the BY50RT simulation produces a maximum helium abundance of $0.35$, which exceeds the observed maximum value of $0.315$ for a GC with the initial mass of $10^6\Msun$.
Therefore, only in the case with a strong magnetic field perpendicular to the cluster's motion does the model pose a significant challenge to the AGB scenario in reproducing the relevant observational constraints on SG fractions and helium abundances in globular clusters. 

Our simulations demonstrate that SG stars exhibit greater central concentration compared to FG stars by the end of our simulation ($t=100\Myr$). We anticipate this spatial configuration to persist relatively unchanged during the long-term evolution of globular clusters, consistent with the majority of observational studies that find SG stars more centrally concentrated in present-day globular clusters \citep{Lardo2011,Simioni2016,Mehta2025}. However, a few observational studies report contrasting results—such as a centrally concentrated FG population or identical radial distributions for multiple populations GCs \citep{Dalessandro2019,Leitinger2023}. These discrepancies may arise from observational challenges \citep{Cadelano2024} that require further investigation. For example, spectroscopic sampling biases in crowded cluster cores could lead to missed SG stars in the central regions \citep{Dondoglio2025}, or long-term dynamical processes might redistribute stellar populations over gigayear timescales. In any case, if these differing distributions of stellar populations in GCs are confirmed, current MSP formation scenarios (particularly the AGB model) would need to incorporate these constraints, as discussed in some studies \citep{Bastian2013C,Howard2019}.
A key finding of our work is the significant influence of magnetic fields and ionizing radiation on shaping SG spatial distributions. Our results show that magnetic fields preferentially confine SG formation to the central regions of the FG cluster, with stronger fields generally producing more compact SG populations - except when the field orientation is perpendicular to the cluster's motion. This magnetic confinement can have important dynamical consequences, as the resulting compact SG configuration appears more resistant to ejection during cluster evolution, thereby enhancing SG retention over time. However, future investigations should particularly focus on long-term dynamical evolution and improved observational constraints on SG distributions in diverse globular cluster environments.

Another aspect we can discuss regarding our results is the perspective of wave propagation in fluid dynamics. When a point source, such as our star cluster, moves with speed $v$ through a medium with sound speed $c_s$, various gas morphologies can arise around the source, depending on the Mach number $M=v/c_s$ \citep{NAKAYAMA2018255}.
If $M<1$ ($v<c_s$), a bow wave forms in front of the cluster. When $M=1$, an underdensity region forms behind the cluster, along with a curved overdensity perpendicular to the source's motion at the cluster's center. In the case of $M>1$ ($v>c_s$), the wavefronts cannot pass through the cluster; instead, they propagate downstream in sequence, creating a cone-shaped shock region known as the Mach cone. \citet{Lai2006} propose a formula for the corresponding Mach angle in a magnetized plasma, indicating that the angle is significantly influenced by the strength and orientation of the magnetic fields. For our moving cluster, the initial Mach number is approximately 15, while the Alfvénic Mach numbers for NoRT simulations with initial magnetic field strengths $0.5$ and $5\mg$ are 60 and 6, respectively. Using that formula, we calculate the Mach angles of the shocked regions for these NoRT runs to be $5$ and $15$ degrees, respectively. These values align very well with the angles observed in our NoRT simulations (see \figref{fig:NoRT}). In the case of
the $50\mg$ magnetic field, the Alfvén velocity is higher than both the cluster's velocity and the sound speed, with the Alfvénic Mach number of about $0.6$. Therefore, the propagation of fast magnetosonic waves is faster along the y-direction than other runs, leading to an overdensity in front of the cluster instead of a cone shape, consistent with the gas morphology observed in the BX50NoRT simulation. While processes such as self-gravity, cooling, star formation, and radiative feedback influence gas properties in the central regions of the cluster, their effects can be neglected in the shocked regions behind the cluster in simulations without radiation. Ionizing radiation significantly alters the gas properties around the cluster, making it challenging to draw similar conclusions for RT runs. With the addition of stellar radiation, the temperature increases, causing the Mach number to decrease to approximately $1.5$ within the cluster, while it falls below unity in the central regions. These changes complicate the explanation of the features observed in RT runs.

While this study provides insights on the effect of magnetic fields on SG formation, it also involves certain simplifications that should be acknowledged. The current work has focused on a constant background magnetic field, neglecting the potential effects of magnetic field turbulence. In astrophysical environments, magnetic fields are likely to be turbulent and variable, which can influence star formation efficiency by affecting gravitational collapse and the expansion of HII regions \citep{Federrath2012,Soam2024}. 
It would be beneficial to conduct similar simulations considering turbulent magnetic fields to further explore the impact of magnetic fields on SG formation. Moreover, we model the proto-globular clusters’ motion using simplified initial conditions, which we acknowledge are uncertain. While such idealized studies have limitations, they provide valuable insights for developing more realistic models of SG star formation. For instance, an inhomogeneous ISM surrounding the cluster could significantly influence the accreted mass. Thus, advancing the AGB scenario requires more detailed modeling of initial conditions within their host galaxy context, taking inspiration from galaxy-scale (and ideally high-resolution cosmological) simulations \citep{Calura2022,Calura2024}.

In this work, we examine the stellar winds retention and pristine gas accretion as key requirements of the AGB scenario to address the abundances problem observed in SG stars. For simplicity, we focus on the He abundance. Our findings indicate that a constant magnetic field does not suppress these processes or the formation of SG stars within a massive  $10^6\Msun$ cluster, except in the case of a very strong $50 \mg$ magnetic field oriented perpendicular to the cluster's motion.  However, the magnetic field can slightly modify the He abundances in the SG cluster, somewhat supporting the AGB scenario to reproduce observations. To determine if this model can explain all observed anomalous chemical compositions, it is essential to explore other light elements, particularly sodium and oxygen. This would require a more comprehensive study in which the yields of individual AGB stars are accurately modeled (Yaghoobi et al., in prep.).

The assumption of a bottom-heavy IMF \citep{bekki2019,Yaghoobi2022a,Yaghoobi2022b} for the SG stars is one of the most significant requirements of the AGB scenario. While the IMF of SG stars remains an unresolved issue, it is generally assumed that the masses of SG stars are typically less than $8 \Msun$ \citep[e.g.,][]{Khalaj2015,Khalaj2016,renzini2015,calura19}. Observational studies show that the stellar mass functions of GCs are predominantly concentrated in a narrow range, primarily featuring low-mass stars \citep[$\le 0.75 \Msun$][]{cadelano20}. 
It has theoretically been demonstrated that the gravitational potential of FG stars can inhibit the formation of more massive stars, effectively truncating the IMF of SG stars \citep{bekki2019}. Previous studies have indicated that the absence of massive stars can occur in systems with an average SFR $\le 10^{-4} \MsunYr$ \citep{Calura2010, Yan2017}. In our simulations, the SFR values are approximately $ \sim 5\times 10^{-4} \MsunYr$, which is slightly higher than that threshold. However, the SFRs associated with SG formation in our simulations are lower than those expected during massive FG formation \citep[e.g., $>10^{-1}\MsunYr$,][]{Polak2024,Cournoyer2024}. In light of such arguments, it might be plausible to expect the number of massive SG stars to be significantly lower than that of massive FG stars. 
Additionally, magnetic fields and ionizing radiation may significantly influence the IMF of SG stars, much like they do for FG stars. Therefore, investigating the IMF of SG stars is crucial for evaluating the AGB scenario in future studies, particularly considering that magnetic fields might have an even greater impact on the primordial IMF than on the present-day IMF. 

\section{Conclusions}\label{sec:Conclusions}
In this work, we study how the combination of magnetic field and photoionization feedback affects the formation of second-generation stars within a young ($34\Myr$) massive ($10^6$) globular cluster, following the AGB scenario. We assume that the cluster supersonically moves through a uniform ISM with gas densities of $10^{-23}\gocm$ and perform a parameter study on the initial magnetic field strength ($0.5, 5, 50 \mg$) of the ISM and its orientation (parallel and perpendicular to the cluster's motion).  Our main results can be listed as:

\begin{itemize}
\item  The pure hydrodynamic simulation reveals that a dense tail forms behind the cluster due to the cluster's supersonic motion, gas accretion, self-gravity, and gas cooling. By adding magnetic fields parallel to the cluster's motion, the propagation of fast magnetosonic waves leads to the extension of this dense tail perpendicular to the magnetic field. These magnetized simulations without ionizing radiation highlight the significant influence of magnetic fields on gas morphology and star-forming regions surrounding a moving massive cluster. We observe a non-linear relationship between the total mass of SG stars, the SG profile, and the maximum He abundance with varying magnetic field strengths.

\item Incorporating photoionizing radiation in parallel magnetic field runs merges the effects of magnetic fields and photoionization feedback. This combination raises gas pressure around the cluster and enhances diffusivity in the intracluster medium due to radiative heating. Consequently, the effectiveness of the magnetic field is somewhat reduced compared to cases without radiation. We find a slight increase in the final SG mass for $0.5$ and $5\mg$ field runs, and a $25 \%$ increase for $50\mg$ case relative to the radiative run without magnetic field. In these cases, magnetic fields continue to limit star-forming regions and SG stars in the cluster's central regions.

\item In radiative runs with magnetic fields oriented perpendicular to the cluster's motion, the gas geometry behind the cluster dissipates more than in parallel cases, due to the impact of stronger magnetic forces. The SG mass remains unchanged for weak and moderate fields, but significantly decreases for the strong  $50\mg$ field.

\item The presence of weak and moderate magnetic fields ($0.5$ and $5\mg$) leads to a slight increase in gas accretion into the cluster's cluster, regardless of orientation. In the case of stronger magnetic fields, notable differences arise: there is a $60$ percent increase (compared to the pure radiative simulation) in accreted pristine material at the center for the parallel field orientation, while a $95$ percent decrease is observed for the perpendicular run.
Moreover, we find that magnetic fields have a negligible effect on central stellar wind retention in simulations with weak magnetic fields and the moderate field simulation with parallel orientation. However, simulations with perpendicular moderate field and strong magnetic fields significantly reduce stellar wind retention.

\item Due to the varying effects of magnetic fields on gas accretion and stellar wind retention, the presence of magnetic fields in our simulations leads to the formation of stars with slightly lower helium abundances — except in the case of strong fields oriented perpendicular to the cluster's motion. This particular run may challenge the AGB scenario to reproduce observations (in terms of He abundance and the mass fraction of SG stars), while the other simulations generally align well with the relevant observational data.

\item Overall, the combined influences of magnetic fields and ionizing radiation are critical in shaping SG clusters. Magnetic fields constrain SG formation in the central regions of the FG cluster, with stronger magnetic fields tending to produce more compact SG clusters, particularly when the field is not oriented perpendicularly to the cluster's motion. 
\end{itemize}

\section*{Acknowledgements}
AY thanks Dr. Reza Rezaei for fruitful discussions. We acknowledge support and computational resources from the PSMN (Pôle Scientifique de Modélisation Numérique) of the ENS de Lyon.

\section*{Data Availability}
The data supporting this study's findings are available from the corresponding author upon reasonable request.

\bibliographystyle{mnras}
\bibliography{example} 

%
%
\appendix
\section{Gas density and Temperature slices for RMHD simulations}\label{Appendix1}
\begin{figure*}
\centering
\includegraphics[width=0.7\linewidth]{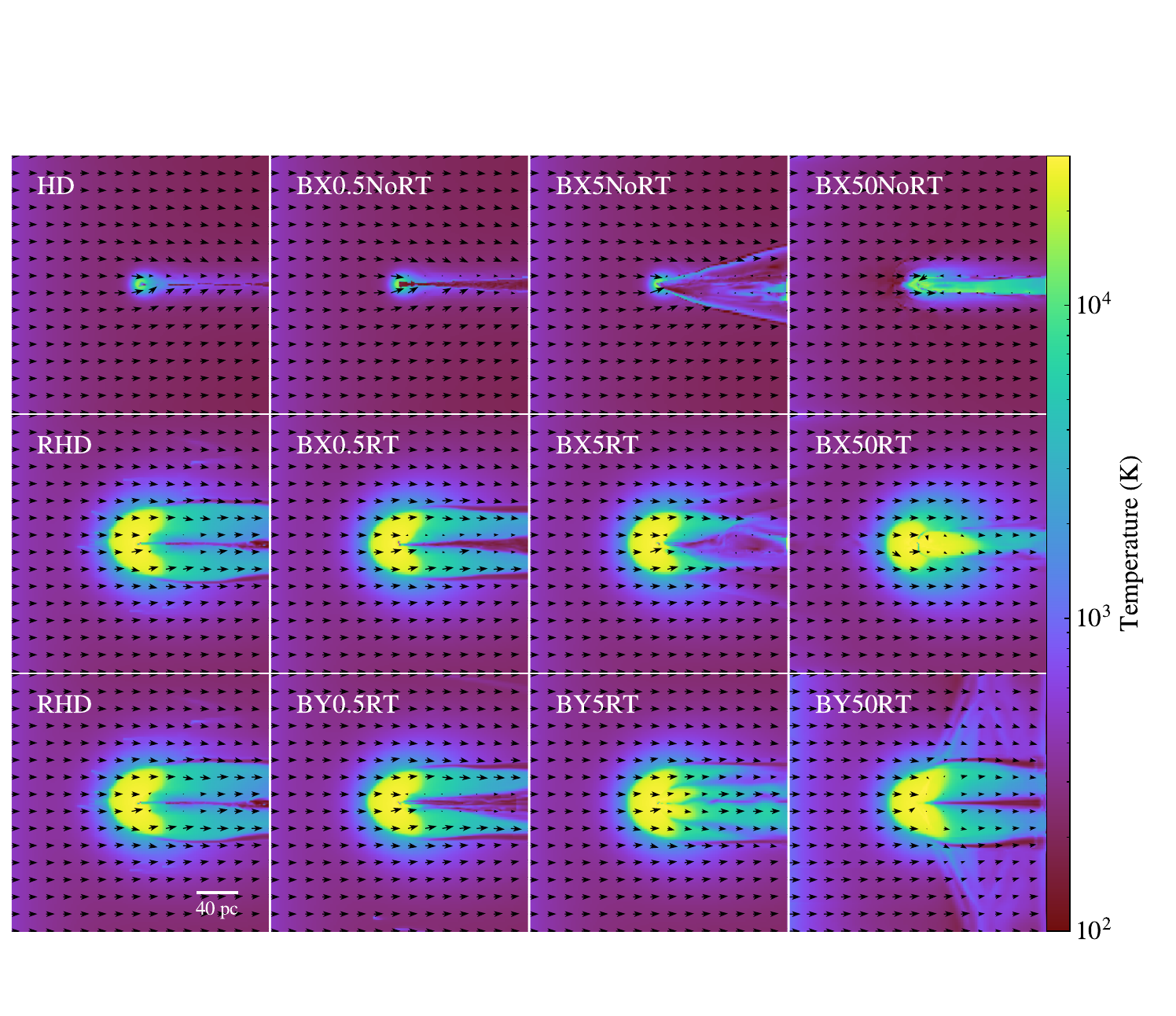}
\caption{Temperature slices computed in the x-y plane for our models at the end of simulations. The black arrows represent the gas velocity field.}	
\label{fig:temp}
\end{figure*}

\figref{fig:temp} shows temperature slices at the end of our simulations, overlaid with arrows representing the velocity fields. The incoming gas with $T=500\K$ enters into the box from the left side and cools to $T\approx200\K$ near the right boundary in the NoRT runs, while in the central regions and behind the cluster —excluding the star-forming regions that reach the temperature floor— the gas reaches $T>400 \K$ due to stellar wind heating (see the top panels of \figref{fig:temp}). In the RT simulations, the inclusion of photoionization feedback heats the gas surrounding the cluster to similar temperatures within a larger radius of approximately $100 \pc$ (the middle and bottom panels of \figref{fig:temp}). Across all temperature panels, the arrows indicate that the stronger the magnetic field, the more the gas tends to move in alignment with the field lines.

\section{Resolution Convergence}\label{Appendix2}
\begin{figure*}
\centering
\includegraphics[width=0.7\linewidth]{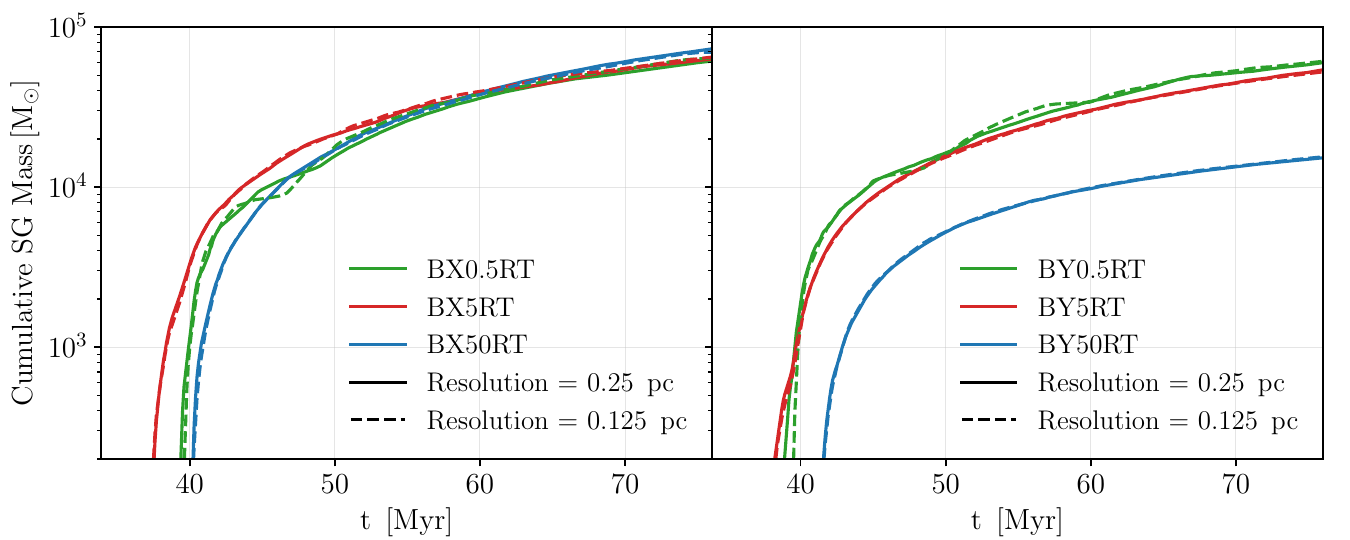}
\caption{The cumulative SG stellar mass as a function of time formed in our RMHD models is presented at different resolutions. The dashed and solid lines represent the results computed at higher ($0.1\pc$) and lower ($0.2\pc$) resolutions, respectively.}	
\label{fig:res}
\end{figure*}

In order to investigate the dependence of the total SG mass on resolution ($\Delta x_{max}$ = $2\pc$ and $\Delta x_{min}$ = $0.25\pc$) in our magnetized simulations, we perform a series of test runs. The results demonstrate that simulations with different resolutions yield approximately the same final results. \figref{fig:res} compares the cumulative SG mass as a function of time for two resolutions of $0.2$ and $0.1\pc$. This indicates that the adopted resolution is sufficient for our results to converge.

\bsp	
\label{lastpage}
\end{document}